------------------------------------------------------------------

\documentstyle[12pt]{article}
\title{{Boson-Realization Model for the}\\
{Vibrational Spectra of Tetrahedral Molecules}}

\author{Zhong-Qi Ma \thanks{email address: MAZQ@BEPC3.IHEP.AC.CN}\\[-2mm]
{\footnotesize CCAST (World Laboratory), PO Box 8730, 
Beijing 100080, and}\\[-2mm]
{\footnotesize Institute of High Energy Physics, 
P.O.Box 918(4), Beijing 100039, P. R. of China}\\
Xi-Wen Hou \\[-2mm]
{\footnotesize Institute of High Energy Physics, 
P.O.Box 918(4), Beijing 100039, and}\\[-2mm]
{\footnotesize Department of Physics, University of Three
Gorges, Yichang 443000, P. R. of China}\\
Mi Xie \\
{\footnotesize Graduate School, Chinese Academy of Sciences, 
Beijing 100039, P. R. of China}}

\date{}

\begin{document}

\maketitle

\vspace{20mm}

\begin{abstract}
An algebraic model of Boson realization is proposed to study the 
vibrational spectra of a tetrahedral molecule, where ten sets of 
boson creation and annihilation operators are used to construct
the Hamiltonian with $T_{d}$ symmetry. There are two schemes
in our model. The first scheme provides an eight-parameter fit 
to the published experimental vibrational eigenvalues of methane 
with a root-mean-square deviation 11.61 $cm^{-1}$. The
second scheme, where the bending oscillators are assumed to be
harmonic and the interactions between the bending vibrations are 
neglected, provides a five-parameter fit with a root-mean-square 
deviation 12.42 $cm^{-1}$.

\end{abstract}

\newpage
\begin{center}
{\bf I. INTRODUCTION}
\end{center}

\vspace{2mm}
The characterization of highly excited vibrational states has 
become one of the central goals in chemical physics. There were 
two general methods used to describe molecular vibrations.
In the traditional approach the molecular Hamiltonian was 
parametrized in terms of internal coordinates ${[1]}$.
The potential was modeled by the force field constants
with many parameters due to poor knowledge of the large
number of force constants. Those parameters have to be 
determined by fitting the spectroscopic data phenomenologically.

As an alternative, an algebraic approach has been proposed for 
the study of polymolecular spectra. The first step toward the 
establishment of an algebraic approach was given by Iachello,
Levine and their co-workers ${[2]}$ with the vibron model, where the 
rotation-vibration spectra of diatomic molecules are described 
in terms of a $u(4)$ algebra. Although this model was extended 
${[3]}$ to polyatomic molecules by introducing a $u(4)$ 
algebra for each bond, it is rather difficult to apply when
the number of atoms in the molecule becomes larger than four
${[4]}$.

Recently, an alternative technique ${[5]}$ for the automatic computation 
of symmetrized local mode basis functions was used to provide
a four parameter potential model for the stretching modes
of octahedron $XY_{6}$ that gave an excellent fit to the published
experimental vibrational eigenvalues of $SF_{6}$, $WF_{6}$ and 
$UF_{6}$. An improved algebraic method ${[6]}$, where the 
one-dimensional Morse oscillator was described by the Lie algebra 
$u(2)$, was proposed to provide another better fit to those 
experimental data with four parameters plus one fixed parameter
$N$ that describes the anharmonic property of the Morse potential. 
This algebraic approach was extended to study the vibrational spectra, 
both stretching and bending, of tetrahedral molecules ${[4]}$,
where the interactions between stretching and bending vibrations
were neglected and seven adjustable parameters plus two fixed
parameters $N_{i}$ were used to fit the experimental data.
This algebraic method was also used to study the vibrational
spectroscopy and intramolecular relaxation of benzene ${[7]}$.

In this paper we propose another algebraic model,
the boson realization model, to study
the vibrational spectra of tetrahedral molecules, where
10 coupled one-dimensional anharmonic oscillators are
described by 10 sets of bosonic creation and annihilation 
operators. The interbond interactions
and the interactions between stretching and bending vibrations
are expressed by the $T_{d}$ invariant combinations of the 
products of one creation operator and one annihilation 
operator such that the total number of vibrational quanta 
is conservative. The symmetrized bases are used to simplify
the calculation. There are two schemes in our model. The 
first scheme provides an eight-parameter fit to the published 
experimental vibrational eigenvalues of methane better
than the previous results.  The results 
show that the bending oscillators are near harmonic ones, the 
interbond interactions between bending vibrations are 
quite weak,  and the interactions between stretching and 
bending vibrations are strong. Those conclusions reflect the 
properties of the molecular structure of methane. From the 
properties we propose our second scheme where the bending oscillators 
are harmonic and there is no interaction between the 
bending vibrations. The second scheme provides a five-parameter 
fit to the experimental data of methane with the root-mean-square 
deviation 12.42 $cm^{-1}$. It may be a model with the least 
parameters that well fits the published experimental 
vibrational data of methane. 

To some extent, our method is a 
generalization of that used in Ref. [5].

This paper is organized as follows. In Sec. II the vibrational 
Hamiltonian of a tetrahedral molecule is introduced in terms of
ten sets of bosonic operators. In Sec. III the vibrational functions 
are combined into the symmetrized bases belonging to given rows 
of given irreducible representations of $T_{d}$. In these 
symmetrized bases the Hamiltonian becomes a block matrix with
eight parameters. The spurious states are ruled out in the 
calculation. In Sec. IV, fitting the published 19 experimental 
data ${[4,8]}$ for methane with the total number of quanta 
$v\leq 3$, in our first scheme with eight parameters we obtain 
the root-mean-square deviation of energy to be 11.61 $cm^{-1}$. 
Our second scheme provides a good five-parameter fit. For comparison, 
the experimental data, the previous calculated energies by the 
algebraic model ${[4]}$ and the present calculated results 
for methane of $v\leq 3$ in two schemes are listed in Table 2. 
The remaining calculated energies for methane up to $v=3$ by this 
boson realization model in the first scheme are also presented. 
The higher energy levels can be calculated straightforwardly. 
In Sec. V we give some conclusions.

\vspace{10mm}
\begin{center}
{\bf II. HAMILTONIAN}
\end{center}

\vspace{2mm}
We begin with enumerating 10 oscillators for an $XY_{4}$ tetrahedral
molecule such as that in Fig. 1. The atom $X$ is located at the center 
$O$ of the tetrahedron, and four atoms $Y$ at its vertices $A$, $B$, 
$C$, and $D$. The coordinate axes $x$, $y$, and $z$ point from $O$ to 
the centers of edges $AC$, $AD$, and $AB$, respectively. The first four 
equivalent oscillators describe the fundamental stretching modes ($A_{1}
\oplus F_{2}$), and the other six equivalent ones describe
the fundamental bending modes. As is well known ${[9]}$, there 
are only five degrees of freedom for the bending vibrations 
($E\oplus F_{2}$), so that the six bending oscillators 
must contain a spurious one. The spurious states related to the
spurious degree of freedom should be ruled out in the later
calculation.

\begin{center}

\setlength{\unitlength}{1.0mm}
\begin{picture}(60,80)(0,-10)

\put(30,65){\line(0,-1){30}}
\put(0,25){\line(3,1){30}}
\put(30,35){\line(1,-3){10}}
\put(30,35){\line(3,-1){30}}
\put(30,35){\vector(-3,2){20}}
\put(30,35){\vector(1,0){12}}
\put(30,35){\vector(3,2){20}}
\put(15,45){\circle*{2}}
\put(35,35){\circle*{2}}
\put(45,45){\circle*{2}}
\put(30,35){\circle*{2}}
\put(9,50){\makebox(0,0)[r]{$z$}}
\put(42,36){\makebox(0,0)[b]{$x$}}
\put(51,50){\makebox(0,0)[l]{$y$}}
\put(30,66){\makebox(0,0)[b]{$A$}}
\put(28,33){\makebox(0,0)[t]{$O$}}
\put(-2,25){\makebox(0,0)[r]{$B$}}
\put(40,4){\makebox(0,0)[t]{$C$}}
\put(62,25){\makebox(0,0)[l]{$D$}}
\put(20,55){\makebox(0,0)[b]{$5$}}
\put(40,55){\makebox(0,0)[b]{$7$}}
\put(29,50){\makebox(0,0)[r]{$1$}}
\put(34,50){\makebox(0,0)[l]{$6$}}
\put(15,31){\makebox(0,0)[b]{$2$}}
\put(49,30){\makebox(0,0)[b]{$4$}}
\put(25,24){\makebox(0,0)[t]{$9$}}
\put(35,15){\makebox(0,0)[r]{$3$}}
\put(19,14){\makebox(0,0)[t]{$10$}}
\put(51,14){\makebox(0,0)[l]{$8$}}

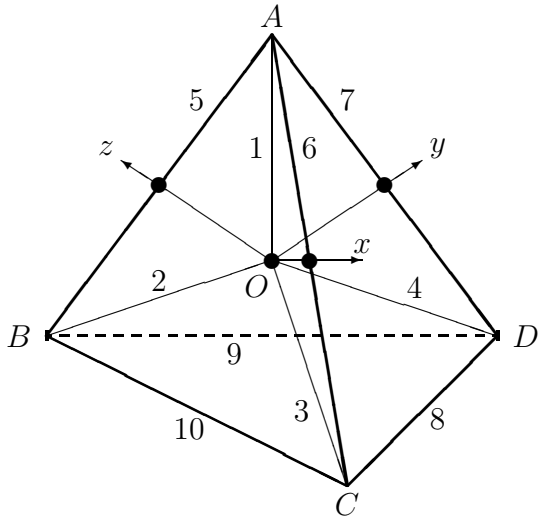
\put(30,-5){\makebox(0,0){{\bæ Fig.1}® Schematic representation of 
a $XY_{4}$ tetrahedral molecule.}}

\thicklines
\put(0,25){\dashbox{1.2}(60,0){}}
\put(0.1,25){\dashbox{1.2}(60,0){}}
\put(0,25){\line(2,-1){40}}
\put(0.1,25){\line(2,-1){40}}
\put(60,25){\line(-1,-1){20}}
\put(60.1,25){\line(-1,-1){20}}
\put(0,25){\line(3,4){30}}
\put(0.1,25){\line(3,4){30}}
\put(30,65){\line(1,-6){10}}
\put(30.1,65){\line(1,-6){10}}
\put(30,65){\line(3,-4){30}}
\put(30.1,65){\line(3,-4){30}}

\end{picture}
\end{center}

Now, for the tenoscillators we introduce 10 sets of bosonic operators 
$a^{\dagger}_{\alpha}$ and $a_{\alpha}$, $1\leq \alpha \leq 10$,
that satisfy the well known relations
$$\begin{array}{l}
[a_{\alpha}~,~a^{\dagger}_{\beta}]~=~\delta_{\alpha \beta},~~~~~
[a_{\alpha}~,~a_{\beta}]
~=~[a^{\dagger}_{\alpha}~,~a^{\dagger}_{\beta}]~=~0, \\ 
|n\rangle~\equiv~|n_{1},n_{2},\cdots,n_{10}\rangle,\\
a^{\dagger}_{\alpha}~|n\rangle
~=~\sqrt{n_{\alpha}+1}~|\cdots,n_{\alpha -1},(n_{\alpha}+1),
n_{\alpha +1},\cdots\rangle ,\\
a_{\alpha}~|n\rangle~=~\sqrt{n_{\alpha}}~
|\cdots,n_{\alpha -1},(n_{\alpha}-1),
n_{\alpha +1},\cdots\rangle,
\end{array} \eqno (2.1) $$

\noindent
where $|n\rangle$ denotes the common eigenstate of the phonon 
number operators $N_{\alpha}$ with the eigenvalues $n_{\alpha}$,
respectively.
$$N_{\alpha}~=~a^{\dagger}_{\alpha}a_{\alpha},~~~~
N_{\alpha}~|n\rangle~=~n_{\alpha}~|n \rangle. \eqno (2.2) $$

The first four bosonic operators $a^{\dagger}_{j}$ (or $a_{j}$), 
$1\leq j \leq 4$, describing the stretching vibrations, 
are the irreducible tensor operators belonging to 
the representations $A_{1}\oplus F_{2}$ of $T_{d}$. The other six  
$a^{\dagger}_{\mu}$ (or $a_{\mu}$), $5 \leq \mu \leq 10$, describing
the bending vibrations, are those 
belonging to the representations $A_{1}\oplus E \oplus F_{2}$. 

The energy of each oscillator depends upon the phonon number. For 
simplicity we assume that all oscillators are the Morse ones 
with two parameters $\omega$ and $x$, so that the energy of the
$\alpha$th oscillator can be expressed in the operator form:
$$\begin{array}{rll}
E_{s}(N_{j})&=~N_{j}~\left\{\omega_{s}~-~x_{s}~(N_{j}+1)\right\},~~~~
&1 \leq j \leq 4\\
E_{b}(N_{\mu})&=~N_{\mu}~\left\{\omega_{b}~-~x_{b}~(N_{\mu}+1)\right\},
&5 \leq \mu \leq 10
\end{array} \eqno (2.3) $$

\noindent
where the subscript $s$ denotes the stretching vibration and $b$ the 
bending one. The null energy has been removed. Although the Morse
potential is known ${[10]}$ to be not very suitable for all anharmonic 
oscillators, the deviation can be described by some more parameters
that become important for higher energy levels. It was pointed out
by Iachello and Oss ${[7]}$ that the P\"{o}schl-Teller potential
is much more appropriate than the Morse one to describe the bending
vibrations. However, the expression for the eigenvalues of the 
bound states is identical for two potentials ${[7]}$.

As usual, neglecting the mixture of the states with different 
total number of phonons, and assuming to take the interactions
up to the second order, we can express the interactive potentials
as the combinations of the products of one creation operator
and one annihilation operator. The character table, the representation 
matrices of the generators, and the Clebsch-Gordan coefficients of the 
$T_{d}$ group were explicitly given in Ref.[4]. From that
knowledge, there are obviously only five $T_{d}$ invariant 
combinations in addition to the sum of phonon number operators. 
The Hamiltonian now can be expressed in terms 
of the bosonic operators as follows:
$$\begin{array}{rl}
H&=~\displaystyle \sum_{j=1}^{4}~E_{s}(a^{\dagger}_{j}a_{j})
~+~\displaystyle \sum_{\mu=5}^{10}~E_{b}(a^{\dagger}_{\mu}a_{\mu})
~+~\lambda_{1}\displaystyle \sum_{i\neq j=1}^{4}~a^{\dagger}_{i}a_{j}
~+~\lambda_{2}\displaystyle \sum_{\mu\neq \nu=5}^{10}~
a^{\dagger}_{\mu}a_{\nu}\\
&~~~~+~\lambda_{3}\displaystyle \sum_{\mu=5}^{7}~\left(
a^{\dagger}_{\mu}a_{\mu +3}+a^{\dagger}_{\mu +3}a_{\mu}\right) 
~+~\lambda_{4}\left\{ \displaystyle 
a^{\dagger}_{1}\sum_{\mu =5}^{7} \left(a_{\mu}-a_{\mu +3}\right)
\right. \\
&~~~~+a^{\dagger}_{2} \left(a_{5}-\displaystyle \sum_{\mu =6}^{8} a_{\mu}
+a_{9}+a_{10}\right)
+a^{\dagger}_{3}\displaystyle \sum_{\mu =3}^{5}
 \left(a_{2\mu}-a_{2\mu -1}\right)\\
&\left.~~~~+a^{\dagger}_{4}\left(-a_{5}-a_{6}+
\displaystyle \sum_{\mu =7}^{9} a_{\mu}
-a_{10}\right)+H.C.\right\}~+~\lambda_{5}\left\{ \displaystyle 
\left(\sum_{j=1}^{4} a^{\dagger}_{j}\right)
 \left(\sum_{\mu=5}^{10} a_{\mu}\right)+H.C. \right\}
\end{array} \eqno (2.4) $$

\noindent
It will be seen in the next section that the term with $\lambda_{5}$
relates only to the spurious states so that it is not interesting
to us. Removing this term, we obtain the Hamiltonian containing
eight parameters. The term with $\lambda_{4}$ describes
the interactions between stretching and bending vibrations. In the 
previous algebraic approach ${[4]}$ ten parameters were introduced 
to fit the experimental data, where two spectroscopic constants 
$N_{1}$ and $N_{2}$, that are equal to $\omega_{s}/x_{s}-1$ and 
$\omega_{b}/x_{b}-1$ in our notations, were taken as fixed parameters, 
one constraint was assumed to reduce one parameter, and
the interactions between stretching and bending vibrations 
were neglected. In this meaning their
algebraic approach ${[4]}$, in comparison with our model, introduced
three more parameters ($B_{12}$, $B_{5,6}$, and $B_{5,10}$),
fixed two parameters $N_{1}$ and $N_{2}$, reduced one parameter 
by a constraint, and
neglected one parameter $\lambda_{4}$ describing the interaction.

\vspace{10mm}
\begin{center}
{\bf III. IRREDUCIBLE BASES}
\end{center}

\vspace{2mm}
Since the Hamiltonian has $T_{d}$ symmetry, each eigenfunction
can be combined such that it belongs to a given row of an irreducible 
representation of $T_{d}$. The states that belong to the same 
irreducible space as partners must correspond to the same energy ${[11]}$.
This degeneracy of the partners is called normal one. A symmetric 
perturbation never splits a normal degeneracy. The calculation for 
energy levels will be greatly simplified if those bases are used. 

The vibrational state of a tetrahedral molecule is described
by the phonon numbers $n_{\alpha}$ of the ten oscillators. The 
first four numbers $n_{j}$ describe the stretching vibrations, and 
the next six numbers $n_{\mu}$ describe the bending vibrations.
Those states can be combined into the irreducible bases belonging
to given rows of given irreducible representations, respectively.
For the general vibrations, those states with both stretching and 
bending vibrations should be further combined.

Now, we discuss the combinations of the states for pure
stretching vibrations. 
Briefly denote $|n_{1} n_{2}n_{3}n_{4}000000\rangle$
by $|abcd\rangle$, where the vanishing $n_{\mu}$ ($\mu\geq 5$) are 
neglected in this notation. Firstly assume that $a$, $b$, $c$ and 
$d$ are all different from each other. Under the transformations of 
$T_{d}$ there are $24$ independent states that span the regular 
representation space of $T_{d}$. Through the standard group theory
method ${[11]}$ they can be combined into orthogonal bases belonging 
to ten irreducible representations: $\phi(A_{1},abcd)$, 
$\phi(A_{2},abcd)$, $\phi_{\nu}^{(1)}(E,abcd)$, 
$\phi_{\nu}^{(2)}(E,abcd)$, $\phi_{\nu}^{(1)}(F_{2},abcd)$, 
$\phi_{\nu}^{(2)}(F_{2},abcd)$, $\phi_{\nu}^{(3)}(F_{2},abcd)$, 
$\phi_{\nu}^{(1)}(F_{1},abcd)$, $\phi_{\nu}^{(2)}(F_{1},abcd)$, 
and $\phi_{\nu}^{(3)}(F_{1},abcd)$. 

Similarly, for pure bending vibrations, briefly denote  
$|0000n_{5} n_{6}n_{7}n_{8}n_{9}n_{10}\rangle$
by $|abcdef\rangle$, where the vanishing $n_{j}$ ($j\leq 4$) are neglected 
in this notation. Two kinds of states should not be confused: one with four 
numbers describes the stretching vibration, and the other with six numbers
describes the bending vibration.
When $a$, $b$, $c$, $d$, $e$ and $f$ are 
all different from each other, we also have $24$ independent states, 
spanning the regular representation space of $T_{d}$. 
Combine them into the irreducible orthogonal bases, and denote them
by $\psi(A_{1},abcdef)$, 
$\psi(A_{2},abcdef)$, $\psi_{\nu}^{(1)}(E,abcdef)$, 
$~\psi_{\nu}^{(2)}(E,abcdef)~$, $~\psi_{\nu}^{(1)}(F_{2},abcdef)~$, 
$\psi_{\nu}^{(2)}(F_{2},abcdef)$, $\psi_{\nu}^{(3)}(F_{2},abcdef)$, 
$\psi_{\nu}^{(1)}(F_{1},abcdef)$, $\psi_{\nu}^{(2)}(F_{1},abcdef)$, 
and $\psi_{\nu}^{(3)}(F_{1},abcdef)$. 
The explicit combinations of $\phi_{\nu}(\Gamma, abcd)$ and
$\psi_{\nu}(\Gamma,abcdef)$ can be obtained from us upon request. 

For example, for the fundamental stretching vibrations ($v=1$) 
the irreducible bases are listed as follows:
$$\begin{array}{rl}
\phi(A_{1},1000)&=~2^{-1} ~\left\{
 |1000 \rangle  ~+~|0100 \rangle  ~+~|0010 \rangle     ~+~|0001 \rangle 
 \right\}\\
\phi_{1}^{(1)}(F_{2},1000)&=~2^{-1}~\left\{
 |1000 \rangle   ~-~|0100 \rangle ~+~|0010 \rangle ~-~|0001 \rangle 
 \right\}\\
\phi_{2}^{(1)}(F_{2},1000)&=~2^{-1}~ \left\{
 |1000 \rangle ~-~|0100 \rangle  ~-~|0100 \rangle  ~+~|0001 \rangle 
\right\}\\
\phi_{3}^{(1)}(F_{2},1000)&=~2^{-1}~\left\{
 |1000 \rangle  ~+~|0100 \rangle ~-~|0010 \rangle  ~-~|0001 \rangle 
 \right\}
\end{array} \eqno (3.1) $$

\noindent
Similarly, the irreducible bases of the fundamental bending 
vibrations ($v=1$) are:
$$\begin{array}{rl}
\psi(A_{1},100000)&=~6^{-1/2} \left\{
 |100000 \rangle  ~+~|010000 \rangle  ~+~|001000 \rangle ~+~|000100 \rangle 
 \right. \\
&\left.~~~~~~~~+~|000010 \rangle  ~+~|000001 \rangle\right\}\\
\psi_{1}^{(1)}(E,100000)&=~(2\sqrt{3})^{-1}~\left\{
 2|100000 \rangle ~-~|010000 \rangle ~-~|001000 \rangle \right. \\ 
&\left.~~~~~~~~~~~+~ 2|000100 \rangle ~-~|000010 \rangle ~-~|000001 \rangle 
 \right\}\\
\psi_{2}^{(1)}(E,100000)&=~2^{-1}~ \left\{
 |010000 \rangle ~-~|001000 \rangle  ~+~|000010 \rangle  ~-~|000001 \rangle 
\right\}\\
\psi_{1}^{(1)}(F_{2},100000)&=~2^{-1/2}~\left\{
 |010000 \rangle   ~-~|000010 \rangle  \right\}\\
\psi_{2}^{(1)}(F_{2},100000)&=~2^{-1/2}~ \left\{
 |001000 \rangle ~-~|000001 \rangle \right\}\\
\psi_{3}^{(1)}(F_{2},100000)&=~2^{-1/2}~\left\{
 |100000 \rangle ~-~|000100 \rangle \right\}
\end{array} \eqno (3.2) $$

In those bases the Hamiltonian $H$ given in (2.4) becomes a block
matrix with the submatrices $H(\Gamma, v)$, where $\Gamma$ denotes 
the irreducible representation, and $v$ is the total phonon number. 
Obviously, $H(E,1)$ is a $1 \times 1$ submatrix, but $H(A_{1},1)$
and $H(F_{2},1)$ are $2 \times 2$ submatrices:
$$\begin{array}{rl}
H(E,1)&=~\omega_{b}-2x_{b}-\lambda_{2}+\lambda_{3} \\
H(A_{1},1)&=~\left(\begin{array}{cc} \omega_{s}-2x_{s}
+3\lambda_{1} & 2\sqrt{6} \lambda_{5}\\
2\sqrt{6} \lambda_{5} & \omega_{b}-2x_{b}
+5\lambda_{2}+\lambda_{3} \end{array} \right) \\
H(F_{2},1)&=~\left(\begin{array}{cc} \omega_{s}-2x_{s}
-\lambda_{1} & 2\sqrt{2} \lambda_{4}\\
2\sqrt{2} \lambda_{4} & \omega_{b}-2x_{b}
-\lambda_{2}-\lambda_{3} \end{array} \right) \end{array} \eqno (3.3) $$

\noindent
Note that the state $\psi(A_{1},100000)$ represents the 
fundamental spurious state and should be ruled out. 

In the traditional approach, the higher excited states
are calculated by symmetrizing the products of the fundamental
vibrational states ${[9]}$. However, it may be more easy
to use the irreducible bases, subtracting 
the spurious states, for calculating the excited states.
In the following, we calculate the excited states for
the case with total phonon number $v=2$ in detail. It is
straightforward to calculate higher excited states in this way.

Before calculation, we have to study an
important problem of how to remove the spurious
states. In the recent papers we see two methods of removal. 
Iachello and Oss ${[7]}$ placed the
spurious states at the energies $\geq 10$ times the energies
of the physical states by the projection operators. This 
method of removal is exact for harmonic bending vibrations 
and acquires a small error for anharmonic
one. Instead, Lemus and Frank ${[4]}$ directly eliminated
the spurious states from both the space and the Hamiltonian.
They demanded the matrix elements of the Hamiltonian related with
the fundamental spurious state $\psi(A_{1},100000)$ vanishing.
In our notation, they introduced a constraint in addition to 
$\lambda_{5}=0$ that was assumed in Ref.[4] at beginning:
$$\omega_{b}-2x_{b}
+5\lambda_{2}+\lambda_{3}~=~0 \eqno (3.4) $$

\noindent
so that $H(A_{1},1)$ in (3.3) contains only one non-vanishing elements.

In the present paper we develop the second method of removal. 
First of all, it seems to us that the constraint (3.4)
is not necessary and reasonable because it restricts only the energy 
of the fundamental spurious state to be vanishing, but the 
energies of all other spurious states are non-vanishing. 
Secondly, we have to answer the problem how to identify 
the spurious states. Generally speaking, a state
is identified as a spurious state if it contains $\psi(A_{1},100000)$ 
as a factor. In Ref. [4] (p.8327) the simply additive definition 
for the product of two functions is used:
$$\begin{array}{l}
|n\rangle~|m\rangle ~=~|(n+m)\rangle \\
|(n+m)\rangle~=~|(n_{1}+m_{1}),(n_{2}+m_{2}),\cdots,(n_{10}+m_{10})\rangle
\end{array} \eqno (3.5) $$

\noindent
When two states $|n\rangle$ and $|m\rangle$ describe a pure stretching 
vibration and a bending vibration, respectively, (3.5) is correct.
However, in the calculations of removing the spurious states, both states
describe bending vibrations where (3.5) may not be suitable. 

Borrowing the idea from Ref.[7], we want to find an identification 
rule for the spurious state such that the spurious species are 
separated, if possible, from the physical species in the matrix 
form of the Hamiltonian. As is well known, in the formalism of 
the boson realization the states $|n\rangle$ contains a factor 
$(n!)^{-1/2}$:
$$|n\rangle~=~(n!)^{-1/2}~\left(a^{\dagger}\right)^{n}~|0\rangle$$

\noindent
Therefore, we embed a factor in the definition (3.5) for product:
$$\begin{array}{l}
|n\rangle~|m\rangle ~=~\displaystyle \prod_{\mu}~\left\{{(n_{\mu}+m_{\mu})!
\over n_{\mu}!m_{\mu}!}\right\}^{1/2}~|(n+m)\rangle \\
|(n+m)\rangle~=~|(n_{1}+m_{1}),(n_{2}+m_{2}),\cdots,(n_{10}+m_{10})\rangle
\end{array} \eqno (3.6) $$

\noindent
In terms of this definition we calculate
the general spurious states and find that the off-diagonal 
elements of the Hamiltonian between the spurious species and the 
physical species linearly depend upon $x_{b}$ and $\lambda_{5}$, 
namely, under the conditions that $x_{b}=0$ and $\lambda_{5}=0$ the 
spurious species are totally separated from the physical species
in the matrix form of the Hamiltonian.

The condition $\lambda_{5}=0$ is acceptable because it only appears
in those off-diagonal elements. The condition $x_{b}=0$ means that 
the bending vibrations are harmonic. Fortunately, to our knowledge, 
in the known results $x_{b}$ is quite small (${\it e.g.}$ see Ref. 
[6], [12] and our results below). It is interesting to notice that 
the first method ${[7]}$ of removal is exact also only for harmonic 
bending vibrations. 

Now, we turn back to calculate the excited states with $v=2$.
When $v=2$, the stretching vibrational states are separated 
into 5 sets: $\phi(A_{1},2000)$, 
$\phi_{\nu}^{(1)}(F_{2},2000)$, $\phi(A_{1},1100)$, 
$\phi_{\nu}^{(1)}(E,1100)$, and $\phi_{\nu}^{(1)}(F_{2},1100)$,
and the bending vibrational states are separated into 10 sets:
$\psi(A_{1},200000)$, $\psi_{\nu}^{(1)}(E,200000)$, 
$\psi_{\nu}^{(1)}(F_{2},200000)$, $\psi(A_{1},100100)$, 
$\psi_{\nu}^{(1)}(E,100100)$, $\psi^{(1)}(A_{1},110000)$, 
$\psi_{\nu}^{(1)}(E,110000)$, $\psi_{\nu}^{(1)}(F_{2},110000)$, 
$\psi_{\nu}^{(3)}(F_{2},110000)$, and  
$\psi_{\nu}^{(1)}(F_{1},110000)$. For the mixture states 
$\Psi_{\nu}(\Gamma \in \Gamma_{1}\otimes \Gamma_{2},v)$
of stretching and bending vibrations with $v=2$, we have to combine
the stretching states $\phi_{\nu}(\Gamma_{1},v=1)$ and the 
bending states $\psi_{\nu}(\Gamma_{2},v=1)$ by the
Clebsch-Gordan coefficients of $T_{d}$. 

In terms of the definition (3.6), direct calculation shows that 
there are 5 sets of spurious states with $v=2$, belonging to the following
irreducible representations: two $A_{1}$, one $E$, and two $F_{2}$.
In the following we list only one spurious state for each irreducible
representation space:
$$\begin{array}{l}
\phi(A_{1},1000)~\psi(A_{1},100000),\\
\left\{\psi(A_{1},200000)+\psi(A_{1},100100)+2\psi(A_{1},110000)\right\}/
\sqrt{3}
~=~\left\{\psi(A_{1},100000)\right\}^{2},\\
\left\{\psi_{2}^{(1)}(E,200000)+\psi_{2}^{(1)}(E,100100)
+\psi_{2}^{(1)}(E,110000)\right\}/\sqrt{3}\\
~~~~~~~~~~~~~~~~~~~~~=~\psi(A_{1},100000) \psi_{2}^{(1)}(E,100000),\\
\phi_{3}^{(1)}(F_{2},1000) \psi(A_{1},100000),\\
\left\{\psi_{3}^{(1)}(F_{2},200000)+\sqrt{2} \psi_{3}^{(1)}(F_{2},110000)
\right\}/\sqrt{3}
~=~\psi(A_{1},100000) \psi_{3}^{(1)}(F_{2},100000).
\end{array} \eqno (3.7) $$

Removing the spurious states, we obtain the physical states
belonging to given irreducible representations. There are
five states belonging to representation $A_{1}$, 
five sets of states belonging to $E$, seven sets belonging to $F_{2}$, 
and three sets belonging to $F_{1}$. Only one state for each irreducible 
representation space is listed in the following:
$$\begin{array}{rl}
f_{1}(A_{1},2)&=~\phi(A_{1},2000),\\
f_{2}(A_{1},2)&=~\phi(A_{1},1100),\\
f_{3}(A_{1},2)&=~\left\{\psi(A_{1},200000)-\psi(A_{1},100100)
\right\}/\sqrt{2},\\
f_{4}(A_{1},2)&=~\left\{\psi(A_{1},200000)+\psi(A_{1},100100)
-\psi(A_{1},110000)\right\}/\sqrt{3},\\
f_{5}(A_{1},2)&=~\displaystyle \sum_{\nu=1}^{3}~
\phi_{\nu}^{(1)}(F_{2},1000) \psi_{\nu}^{(1)}(F_{2},100000) /\sqrt{3},
\end{array} $$
$$\begin{array}{rl}
f_{1}(E,2)&=~\phi_{2}^{(1)}(E,1100),\\
f_{2}(E,2)&=~\left\{\psi_{2}^{(1)}(E,200000)-\psi_{2}^{(1)}(E,100100)
\right\}/\sqrt{2},\\
f_{3}(E,2)&=~\left\{\psi_{2}^{(1)}(E,200000)+\psi_{2}^{(1)}(E,100100)
-2\psi_{2}^{(1)}(E,110000)\right\}/\sqrt{6},\\
f_{4}(E,2)&=~\phi(A_{1},1000) \psi_{2}^{(1)}(E,100000),\\
f_{5}(E,2)&=~\left\{\phi_{1}^{(1)}(F_{2},1000)\psi_{1}^{(1)}(F_{2},100000)
-\phi_{2}^{(1)}(F_{2},1000)\psi_{2}^{(1)}(F_{2},100000) \right\}/\sqrt{2},
\end{array} $$
$$\begin{array}{rl}
f_{1}(F_{2},2)&=~\phi_{3}^{(1)}(F_{2},2000),\\
f_{2}(F_{2},2)&=~\phi_{3}^{(1)}(F_{2},1100),\\
f_{3}(F_{2},2)&=~\left\{\sqrt{2}\psi_{3}^{(1)}(F_{2},200000)
-\psi_{3}^{(1)}(F_{2},110000)\right\}/\sqrt{3},\\
f_{4}(F_{2},2)&=~\psi_{3}^{(3)}(F_{2},110000),\\
f_{5}(F_{2},2)&=~\phi_{3}^{(1)}(F_{2},1000) \psi_{1}^{(1)}(E,100000),\\
f_{6}(F_{2},2)&=~\phi(A_{1},1000) \psi_{3}^{(1)}(F_{2},100000),\\
f_{7}(F_{2},2)&=~\left\{\phi_{1}^{(1)}(F_{2},1000)
\psi_{2}^{(1)}(F_{2},100000)+\phi_{2}^{(1)}(F_{2},1000) 
\psi_{1}^{(1)}(F_{2},100000)\right\}/\sqrt{2} ,
\end{array} $$
$$\begin{array}{rl}
f_{1}(F_{1},2)&=~\phi_{3}^{(1)}(F_{1},110000),\\
f_{2}(F_{1},2)&=~\phi_{3}^{(1)}(F_{1},1000)
\psi_{2}^{(1)}(E,100000),\\
f_{3}(F_{1},2)&=~\left\{\phi_{1}^{(1)}(F_{2},1000)
\psi_{2}^{(1)}(F_{2},100000)-\phi_{2}^{(1)}(F_{2},1000) 
\psi_{1}^{(1)}(F_{2},100000)\right\}/\sqrt{2},
\end{array} $$

\noindent
where the number 2 in the argument denotes $v=2$.
Those states belonging to the same irreducible representation 
will be mixed by the Hamiltonian. 

Directly calculating the Hamiltonian matrix $H$ in 
those bases, we obtain a block matrix with $5\times 5$ submatrix
for $A_{1}$, $5\times 5$ submatrix for $E$, $7\times 7$ submatrix
for $F_{2}$, and $3\times 3$ submatrix for $F_{1}$:
$$\begin{array}{l}
{\small H(A_{1},2)~=~
\left(\begin{array}{ccccc}
2\omega_{s}-6x_{s} &2\sqrt{3}\lambda_{1}
 &0&0&2\sqrt{3}\lambda_{4}\\
2\sqrt{3}\lambda_{1} &2\omega_{s}-4x_{s}+4\lambda_{1}
 &0&0&-2\lambda_{4}\\
0&0 &C_{1}&-\sqrt{2/3} x_{b}  &4\lambda_{4}\\
0&0 &-\sqrt{2/3} x_{b}&C_{2}&0\\
2\sqrt{3}\lambda_{4} 
&-2\lambda_{4} &4\lambda_{4} 
&0 &C_{3}
\end{array} \right) }\end{array}$$
$$\begin{array}{l}
{\small H(E,2)~=~
\left(\begin{array}{ccccc}
2\omega_{s}-4x_{s}-2\lambda_{1}
 &0 &0 &0 &4\lambda_{4}\\
0 &C_{1} &-x_{b}/\sqrt{3}
&0 &4 \lambda_{4}\\
0&-x_{b}/\sqrt{3} &C_{4}&0 &0\\
0&0&0 &C_{5}+2\lambda_{3}&0 \\
4\lambda_{4} &4\lambda_{4} 
&0&0 &C_{3}
\end{array}  \right) } \end{array}$$
$$\begin{array}{l}
{\small H(F_{2},2)~=~
\left(\begin{array}{ccccccc}
2\omega_{s}-6x_{s} &2\lambda_{1}  &0&0&0 &2\lambda_{4} &2\sqrt{2}\lambda_{4}\\
2\lambda_{1} &2\omega_{s}-4x_{s} &0&0&0 &2\lambda_{4} &-2\sqrt{2}\lambda_{4}\\
0&0 &C_{6} &0 &2\sqrt{2}\lambda_{4}&0  &0\\  
0&0&0 &C_{7} &0 &0  &4\lambda_{4}\\
0&0 &2\sqrt{2}\lambda_{4} &0 &C_{3}+2\lambda_{3} &0&0 \\
2\lambda_{4}&2\lambda_{4} &0&0&0&C_{5} &0\\
2\sqrt{2}\lambda_{4} &-2\sqrt{2}\lambda_{4} &0 &4\lambda_{4} &0&0 
& C_{3} 
\end{array}  \right) } \end{array}$$
$$\begin{array}{l}
{\small H(F_{1},2)~=~
\left(\begin{array}{ccc}
2\omega_{b}-4x_{b}-2\lambda_{2}
 &-2\sqrt{2}\lambda_{4} &0 \\
-2\sqrt{2}\lambda_{4} &C_{3}+2\lambda_{3} &0\\
0&0 &C_{3}\end{array}  \right) } \end{array}$$

\noindent
where $C_{1}=2\omega_{b}-5x_{b}-2\lambda_{2}-2\lambda_{3}$, 
$C_{2}=2\omega_{b}-(14/3)x_{b}-2\lambda_{2}
+2\lambda_{3}$, 
$C_{3}=\omega_{s}-2x_{s}+\omega_{b}-2x_{b}-\lambda_{1}
-\lambda_{2}-\lambda_{3}$, 
$C_{4}=2\omega_{b}-(13/3)x_{b}-2\lambda_{2}+2\lambda_{3}$, 
$C_{5}=\omega_{s}-2x_{s}+\omega_{b}-2x_{b}+3\lambda_{1}
-\lambda_{2}-\lambda_{3}$, 
$C_{6}=2\omega_{b}-(16/3)x_{b}-2\lambda_{2}$, and
$C_{7}=2\omega_{b}-4x_{b}-2\lambda_{2}-2\lambda_{3}$. 

\vspace{10mm}
\begin{center}
{\bf IV. PURE VIBRATIONAL SPECTRA}
\end{center}

\vspace{2mm}
Now, we are going to fit the experimental data by our 
boson realization model. Methane is a typical molecule with 
$T_{d}$ symmetry. To our knowledge, 
there are 4 data for $v=1$, 7 data for $v=2$ and 8 data for $v=3$.
In our first scheme  we fit those 19 data to determine the eight 
parameters (Table 1, the first scheme) with the root-mean-square energy 
deviation $11.61cm^{-1}$, where the standard deviation is calculated 
unweightedly:
$$\sigma^{2}~=~\displaystyle {1 \over 19-8}~\sum_{i}^{19}~
\left( \nu_{i}({\rm calc})~-~\nu_{i}({\rm expt})\right)^{2} \eqno (4.1) $$

From the results we come to three conclusions:

i) Since $x_{b}$ is relatively small, the bending oscillators 
are near harmonic ones.

ii) The interbond interactions between bending vibrations 
are quite weak.

iii) The interactions between stretching and bending vibrations are
strong. 

From these conclusions, we propose our second scheme where 
the bending oscillators are harmonic ($x_{b}=0$) and
there is no interaction between the bending vibrations 
($\lambda_{2}=\lambda_{3}=0$). The second
scheme provides a five-parameter fit to the experimental data
of methane with the root-mean-square deviation 12.42$cm^{-1}$
(see Table 1, the second scheme). Recall that Ref. [4] presented a 
seven-parameter fit with the root-mean-square deviation 12.16$cm^{-1}$

For comparison, we list in Table 2 the 19 experimental data, the
calculation results from the algebraic model ${[4]}$,
and our results in two schemes for the vibrational spectra 
($v\leq 3$) of methane.

\begin{center}
{\small Table I. Parameters in the Hamiltonian obtained by the least 
square fitting ($cm^{-1}$)}

\vspace{3mm}
\begin{tabular}{|c|c|c|c|c|}
\hline
Scheme& Stretching&Bending&Interac. &\\ \cline{2-5}
&$\omega_{s}~~~~~~~x_{s}~~~~~~~\lambda_{1}$
&$\omega_{b}~~~~~~~~~x_{b}~~~~~~~\lambda_{2}~~~~~\lambda_{3}$ 
&$\lambda_{4}$ &rms \\
\hline
First&2986.74~~77.96~~~34.55 & 1508.37~~-6.635~~-5.96~~-0.90
&-203.73&11.61 \\
\hline
Second&2986.24~~76.55~~~33.60& 1525.85~~~~~~~~~~~~~~~~~~~~~~~~~~~
&-201.65&12.42 \\
\hline
\end{tabular}
\end{center}

\begin{center}
{\small Table II. Experimental data ${[8]}$, algebraic calculation
${[4]}$, and our calculation\\results for the vibration spectra 
($v\leq 3$) of methane ($cm^{-1}$)}.

\vspace{3mm}
\begin{tabular}{|c|c|c|c|c|}
\hline
\multicolumn{5}{|c|}{$v=1$}\\
\hline\hline
$\Gamma$&$F_{2}$&$E$&$A_{1}$&$F_{2}$\\
\hline
Expt. ${[8]}$ &1310.0&1533.0&2916.5&3019.4\\
\hline
Calc. ${[4]}$ &1303.7 &1520.4 &2918.4 &3027.2 \\
\hline
Scheme 1 &1305.7 &1526.7 &2934.5 &3019.0 \\
\hline
Scheme 2 &1307.8 &1525.9 &2933.9 &3017.6 \\
\hline
\end{tabular}

\vspace{3mm}
\begin{tabular}{|c|c|c|c|c|c|c|c|}
\hline
\multicolumn{8}{|c|}{$v=2$ and $F_{2}$}\\
\hline\hline
Expt. ${[8]}$ &2614.0 &2830.4 &4223.0 &4319.0 &4549.0 &5861.0 &6004.7\\
\hline
Calc. ${[4]}$ &2610.5 &2841.5 &4222.0 &4330.9 &4547.7 &5856.7 &6014.5\\
\hline
Scheme 1 &2610.1 &2840.1 &4226.7 &4308.4 &4546.9 &5855.8 &6011.2 \\
\hline
Scheme 2 &2614.3 &2833.6 &4228.8 &4309.5 &4543.5 &5855.3 &6008.6 \\
\hline
\end{tabular}

\vspace{3mm}
\begin{tabular}{|c|c|c|c|c|c|c|c|c|}
\hline
\multicolumn{9}{|c|}{$v=3$ and $F_{2}$}\\
\hline\hline
Expt. ${[8]}$ &4123.0 &5775.0 &5861.0 &7514.0 &8604.0 &8807.0 &8900.0 &9045.0\\
\hline
Calc. ${[4]}$ &4123.9 &5759.9 &5868.7 &7534.9 &8603.0 &8794.1 &8910.0 &9034.5\\
\hline
Scheme 1 &4136.9 &5759.3 &5858.4 &7513.8 &8601.9 &8805.3 &8915.9 &9035.7\\
\hline
Scheme 2 &4140.1 &5754.6 &5851.2 &7534.4 &8603.3 &8804.1 &8913.0 &9031.4\\
\hline
\end{tabular}

\end{center}

In terms of the eight parameters in the first scheme  or the five 
parameters in the second scheme, it is straightforward
to calculate the rest of the vibrational spectra for
methane. After removing the spurious states, for $v=2$,
there are 5 states with $A_{1}$, 5 sets of states with
$E$, 7 sets of states with $F_{2}$, and 3 sets of states with 
$F_{1}$. For $v=3$, there are
13 states with $A_{1}$, 4 states with $A_{2}$, 14 sets of states
with $E$, 25 sets of states with $F_{2}$, and 15 sets of
states with $F_{1}$. Except for the states with $F_{2}$ and $v=2$
that were listed in Table II, the rest of calculation results in 
the first scheme are listed as follows.
$$\begin{array}{llllllll}
v=2,~A_{1} &2614.3 &3057.9 &4297.7 &5807.6 &5974.4&&\\
v=2,~E &2616.5 &3055.6 &4326.3 &4461.2 &6038.2 &&\\
v=2,~F_{1} &2832.4 &4324.8 &4545.7 &&&&\\
\end{array}$$
$$\begin{array}{llllllll}
v=3,~A_{1} &3913.0 &4158.5 &4586.8 &5514.9 &5601.8 &5861.8 &5992.3 \\
&7100.9 &7295.4 &7567.2 &8587.8 &8749.0 &8994.0 &\\
v=3,~A_{2} &4141.8 &4588.6 &5852.6 &7564.8 &&&\\
v=3,~E &4142.0 &4157.5 &4591.3 &5526.7 &5621.8 &5828.1 &5858.1\\
 &5990.1 &7160.1 &7299.4 &7334.5 &7502.1 &7566.1 &8838.7 \\
\end{array}$$
$$\begin{array}{llllllll}
v=3,~F_{2} &3917.6 &3930.3 &4136.9 &4365.5 &4378.3 &5514.3 &5589.1\\
 &5619.9 &5633.9 &5759.3 &5831.1 &5858.4 &6075.8 &6078.6 \\
&7074.8 &7134.5 &7255.9 &7300.1 &7331.0 &7379.4 &7513.8 \\
&8601.9 &8805.3 &8915.9 &9035.7 \\
v=3,~F_{1} &3919.4 &4148.3 &4369.1 &5605.1 &5632.4 &5753.6 &5840.8\\
 &5854.3 &6075.8 &7158.7 &7291.3 &7334.8 &7383.0 &7539.3\\
 &8941.9
\end{array} $$

\vspace{10mm}
\begin{center}
{\bf V. CONCLUSIONS}
\end{center}

\vspace{2mm}
In this paper we describe ten coupled one-dimensional anharmonic 
oscillators of a tetrahedral molecule by ten sets of bosonic 
creation and annihilation operators. The ten oscillators are
divided into two classes: stretching and bending oscillators.
The energy levels of those oscillators are described by four 
parameters under the assumption of the Morse potential for
stretching vibration and the P\"{o}schl-Teller potential for the 
bending vibrations: 
$\omega_{s}$, $x_{s}$, $\omega_{b}$, and $x_{b}$. The interbond 
interactions and the interactions between stretching 
and bending vibrations are supposed to be $T_{d}$ invariant and 
to preserve the total number $v$ of vibrational quanta so that 
4 parameters $\lambda_{i}$, $1\leq i \leq 4$, have to be introduced.

In the first scheme of the boson realization model with eight 
parameters we fit the 19 experimental vibrational data for methane,  
and obtain the root-mean-square energy deviation to be 11.61 $cm^{-1}$.
From the obtained parameters, we see that the interactions between
the bending vibrations are weak, the interactions between
the stretching and bending vibrations are strong, and the
bending oscillators are quite near harmonic ones. These conclusions
are different from the previous model ${[4]}$. From these 
conclusions we proposed another five-parameter fit in the 
second scheme with the root-mean-square energy deviation 12.42$cm^{-1}$.
To our knowledge, it may be the model with the least parameters
that well fits the experimental vibration spectra ($v\leq 3$) of methane.

The interaction between vibrational and rotational motions plays
an important role in describing the abundant experimental data
of vibrorotational energy spectra of a tetrahedral molecule. We will
study it by the boson realization model elsewhere.

\vspace{10mm}
{\bf Acknowledgments}. This work was supported by the National
Natural Science Foundation of China and Grant No. LWTZ-1298 of
the Chinese Academy of Sciences.

\end{document}